\documentclass[letterpaper,10pt,nofootinbib,aps,tightenlines,prd,twocolumn,longbibliography,preprintnumbers]{revtex4-1}
\usepackage{amsmath,amsfonts,amssymb}
\usepackage{mathrsfs} 
\usepackage{graphicx}
\usepackage{color}
\usepackage[english]{babel}
\usepackage[utf8]{inputenc}
\usepackage[sort&compress]{natbib}
\usepackage{soul}

\begin{document}
\title{Cosmic Time Slip: Testing Gravity on Supergalactic Scales with Strong-Lensing Time Delays}
\author{Dhrubo Jyoti${}^1$}
\author{Julian B. Mu\~noz${}^2$}
\author{Robert R. Caldwell${}^1$}
\email{Robert.R.Caldwell@dartmouth.edu}
\author{Marc Kamionkowski${}^3$}
\affiliation{${}^1$Department of Physics \& Astronomy, Dartmouth College, Hanover, NH 03755 USA \\
${}^2$Department of Physics, Harvard University, Cambridge, MA 02138 USA \\
${}^3$Department of Physics \& Astronomy, Johns Hopkins University, Baltimore, MD 21218 USA}

\date{\today}

\begin{abstract}

We devise a test of nonlinear departures from general relativity (GR) using time delays in strong gravitational lenses. 
We use a phenomenological model of gravitational screening as a step discontinuity in the measure of curvature per unit mass, at a radius $\Lambda$.
The resulting slip between two scalar gravitational potentials leads to a shift in the apparent positions and time delays of lensed sources, relative to the GR predictions, of size $\gamma_{\rm PN}-1$. 
As a proof of principle, we use measurements of two lenses, RXJ1121-1231 and B1608+656, to constrain deviations from GR to be below $|\gamma_{\rm PN}-1| \leq 0.2 \times (\Lambda/100\, \rm kpc)$.
These constraints are complementary to other current probes, and are the tightest  in the range $\Lambda=10-200$ kpc, showing that future measurements of strong-lensing time delays have great promise to seek departures from general relativity on kpc-Mpc scales.
\end{abstract}

\maketitle

\section{Introduction}

One of the most perplexing problems of cosmology is to determine the physics of the accelerated cosmic expansion~\cite{Riess:1998cb,Perlmutter:1998np}. While a cosmological constant is widely regarded as the default hypothesis, both dynamical dark energy and new gravitational physics have been put forward as possible explanations~\cite{Weinberg:1988cp,Huterer:2017buf,Caldwell:2009ix,Joyce:2016vqv}. To evaluate the consequences of the myriad such scenarios requires forward modeling the detailed behavior of new fields and interactions. A more nimble comparison of theory with observations could be made if a phenomenological description was available. In the case of dark energy, the commonly used equation of state carries the equivalent information of a quintessence scalar field. In the case of new gravitational physics, efforts have focused on building a cosmological version of the post-Newtonian parametrization (see, for instance, Refs.~\cite{Clifton:2011jh,Koyama:2015vza,Avilez-Lopez:2015dja} for recent reviews). However, there is no one-size-fits-all description of cosmological gravitation beyond general relativity (GR).

Yet, there are two key, distinguishing features of most theories of new gravitational physics: {\it gravitational slip} (meaning different Newtonian potentials for the temporal and spatial metric components~\cite{Daniel:2008et}) that grows with the accelerated cosmic expansion, and {\it screening}, sometimes referred to as Vainshtein screening \cite{Vainshtein:1972sx,Babichev:2013usa}, that maintains GR within the confines of a galaxy, but enables new, light gravitational degrees of freedom to activate on cluster scales and beyond. These features are common to $f(R)$ gravity~\cite{Sotiriou:2008rp}, chameleon fields~\cite{Khoury:2003rn}, beyond-Horndeski gravitation~\cite{Kobayashi:2014ida,Zumalacarregui:2016pph}, and more broadly to theories of massive gravity~\cite{deRham:2014zqa}. Each of these cases require detailed and model-specific calculations to evaluate the predictions for cosmology. We are therefore motivated to posit a phenomenological model (akin to Ref.~\cite{Platscher:2018voh}), in which the departures from GR take the form of a gravitational slip on distances above some cutoff scale $\Lambda$, which is expected to match the general behavior of these complete theories. The cosmological effects of gravitational slip and screening have been studied in a wide range of contexts: cosmic microwave background (CMB) temperature and polarization anisotropies~\cite{Daniel:2010ky,Amendola:2014wma,Munshi:2014tua,Ade:2015rim,Namikawa:2018erh}, weak gravitational lensing~\cite{DAmico:2016ntq,Paillas:2018wxs}, the growth and clustering of large scale structure~\cite{Daniel:2010ky,Nesseris:2017vor,Salzano:2017qac}, strong gravitational lensing~\cite{Smith:2009fn,Bolton:2006yz,Collett:2018gpf,Pizzuti:2016ouw}, and in stars and galaxies \cite{Hui:2012jb,Koyama:2015oma}.

In this paper we propose strong-lensing time delays as a probe of gravitational slip. 
In usual time-delay cosmography one uses the positions and fluxes of a multiply imaged quasar to constrain the lens-mass model. 
In these cases, the strong-lens galaxy (usually a massive elliptical) is in the line of sight between a quasar and us, resulting in multiple images for the quasar.
This leaves the time delay between images as an additional degree of freedom, which can be used to measure the cosmic expansion rate (see, for example, Ref.~\cite{Treu:2016ljm} for a recent review). Using this technique, the H0LiCOW ($H_0$ Lenses in COsmograil Wellspring) collaboration has recently reported a measurement of the Hubble constant of $H_0=72.5^{+2.1}_{-2.3}$ km s$^{-1}$ Mpc$^{-1}$~\cite{Birrer:2018vtm,Bonvin:2016crt}, using four strongly lensed systems, showcasing the strength of time-delay measurements. Here, instead, we propose fixing $H_0$ to its CMB- or supernova-inferred value~\cite{Riess:2016jrr}, and using the time-delay measurements as a test of deviations from GR. Our procedure is relatively straightforward, and fits within the standard framework used to model strong-lensing time delays, which would allow for departures from GR to be constrained by future analyses.

We make two simplifying assumptions in this work. First, we only consider the spherically symmetric images in each lens, though our method can be generalized to use the fully non-spherical information and any substructure. Second, we assume that the screening length $\Lambda$ is bigger than the Einstein radius of the lens galaxy, and therefore significantly larger than its half-light radius. Thus, the stellar dynamics within the lens are not altered, but deviations from GR at large radii would affect the photon time travel.
This is to be compared with the results such as Refs.~\cite{Smith:2009fn,Bolton:2006yz,Collett:2018gpf,Cao:2017nnq,Hossenfelder:2018iym}, where the screening is assumed to take place within the galaxy, and departures from GR are constrained by comparing the dynamical and lensing masses.
We are able to explore the opposite regime of supergalactic screening due to the inclusion of the time-delay datum.
Additionally, previous studies have used strong-lensing time delays as a probe of modified gravity or dark energy \cite{Ishak:2008ex,Coe:2009wt,Paraficz:2009xj,Suyu:2013kha,Treu:2013rpx,Jee:2015yra,Linder:2016bso}, focusing on the changes to the expansion history of the Universe. 
Our work is different from those studies in that it seeks changes to the space-time around the lens rather than to the expansion history of the Universe.

In this first study we use data of real quadruply lensed quasars from the H0LiCOW collaboration, for which the amount of information about the lens is maximal~\cite{Suyu:2016qxx}. 
We show that the data of two lensed systems is already sufficient to obtain new bounds on departures from GR. Indeed, with time-delay measurements of RXJ1131-1231 and B1608+656 we are able to constrain a deviation to the Post-Newtonian slip parameter $|\gamma_{\rm PN}-1| \leq 0.2 \times (\Lambda/100\,\rm kpc)$, 
which sets the most stringent constraints on new theories of gravity with screening lengths $\Lambda=10-200$ kpc.
This technique opens up a new way to probe gravitational phenomena on cosmological scales, where dark-energy effects are expected to become apparent.


We structure this paper as follows.
We introduce our model in Sec.~\ref{sec:Model}, which we use to obtain the time delays in Sec.~\ref{sec:TimeDelays}.
We, then, present our results in Sec.~\ref{sec:Results} and conclude in Sec.~\ref{sec:Conclusions}.

\section{The Model}
\label{sec:Model}

We consider the geodesic motion of photons under a metric theory of gravity in which our cosmological spacetime is described by the line-element
\begin{equation}
    ds^2 = a^2(\eta)\left[-(1 + 2 \Phi) d\eta^2 + (1 - 2 \Psi)d \vec x^2\right].
\end{equation}
Here $a$ is the expansion scale factor, $\eta$ is conformal time, and $\Phi$, $\Psi$ are the conformal-Newtonian and longitudinal potentials, respectively. In the Newtonian limit, valid for length and time scales shorter than the expansion time, a non-relativistic distribution of matter gives rise to a weak potential $|U| \ll 1$, according to the Poisson equation: $\nabla^2 U = 4 \pi G a^2 \rho$. In this case, the acceleration of massive test particles is determined as $\vec x'' = -\vec \nabla \Phi$ with $\Phi = U$. These potentials are equal, $\Psi = \Phi$, under GR~\cite{1992grle.book.....S, Ma:1995ey}.
 
Gravitational slip describes the decoupling of $\Phi$ and $\Psi$ as a consequence of a departure from GR. In the class of models considered, new gravitational degrees of freedom yield $\Psi = \gamma_{\rm PN} \Phi$, where $\gamma_{\rm PN}$ quantifies the amount of space-curvature per unit rest mass, and is expected to return to its GR value of $\gamma_{\rm PN}=1$ at small distances due to screening.

Gravitational screening is a nonlinear phenomenon whereby the same new gravitational degrees of freedom are sharply suppressed within a certain region. 
The simplest theory in which this appears is the cubic galileon \cite{Nicolis:2008in,Deffayet:2009wt,Deffayet:2009mn}, wherein the screening radius is determined by a geometric mean of the Schwarzschild radius of the mass source and the Compton wavelength of the new degrees of freedom. This elegant effect enables these gravitational theories to closely resemble GR within our galaxy, where classical tests strongly favor Einstein's theory, but allows new effects---in particular cosmic acceleration---to manifest on larger scales. To model the effect of screening, we consider the gravitational slip to be stepwise discontinuous at a screening radius $\Lambda$.

 
\begin{figure*}[ht]
\includegraphics[scale=0.3]{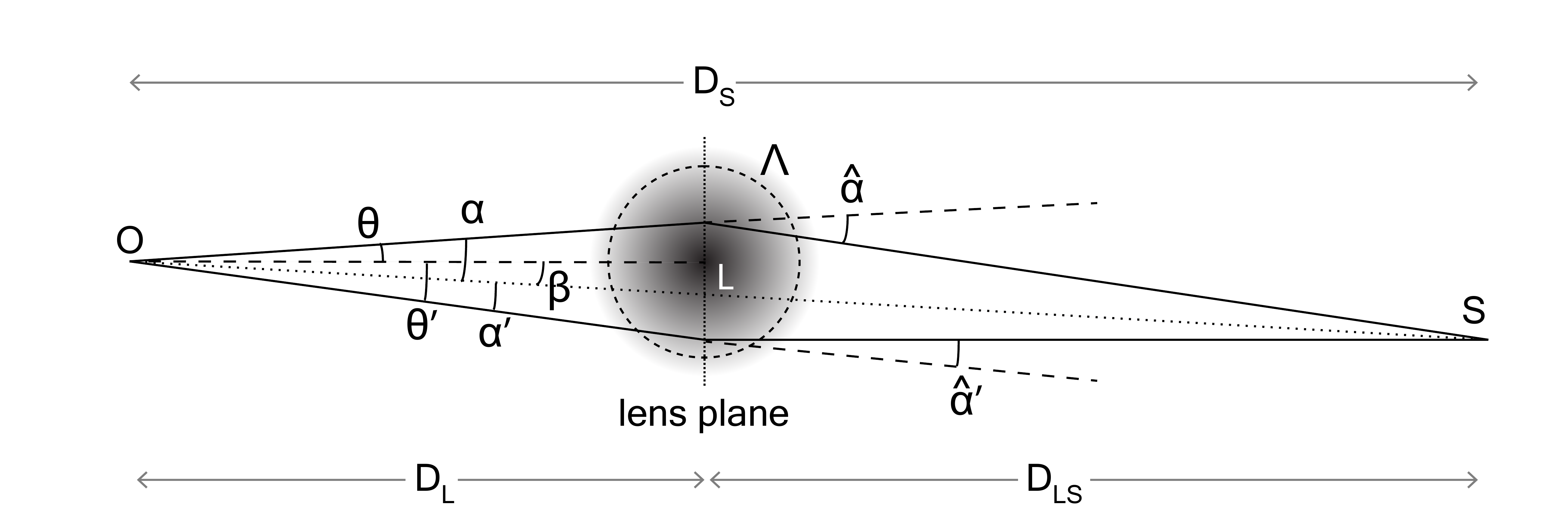} \caption{The relative positions of observer (O), lens (L), and source (S). The observer records two images, displaced by angles $\theta,\,\theta'$ relative to the lens. We assume that the screening radius $\Lambda$, measured from the center of the lens $L$, is larger than the impact parameter of both light rays. 
In our toy model the post-Newtonian parameter $\gamma_{PN}$ is unity inside $\Lambda$, and a free parameter outside, see  Eq.~\eqref{eq:model}. Note that $\hat{\alpha}$ and $\hat{\alpha}'$ are deflection angles, and $\alpha$ and $\alpha'$ are {\it reduced} deflection angles. They are related as follows: $\alpha=(D_{LS}/D_S)\,\hat{\alpha}$~~\cite{Narayan:1996ba}.}
\label{fig:thin_lens}
\end{figure*}


Photon geodesics require the sum of the two potentials, which we define as $\Sigma \equiv \Phi + \Psi$. For a spherically symmetric mass distribution, $\rho(r)$, we propose to model a departure from general relativity as
\begin{equation}
    \Sigma = [2 + (\gamma_{\rm PN}-1)\Theta(r-\Lambda)]\Phi(r),
    \label{eq:model}
\end{equation}
where $r$ and $\Lambda$ are physical distances, and $\Theta$ is the Heaviside step function.
This simple expression is our main innovation, which enables an easy calculation of lensing deflection and time delay, given a spherical lens mass model. Our setup is illustrated in Figure \ref{fig:thin_lens}.  
 In what follows, we will assume the screening radius is larger than the Einstein radius, $\Lambda>R_E=D_L\theta_E$.

We define, as usual, the lensing potential as~\cite{Narayan:1996ba}
\begin{equation}
\psi(\mathbf \theta) = \frac{D_{LS}}{D_L D_S} \int dz\,\Sigma(r),
\end{equation}
where $D_L$, $D_S$ and $D_{LS}$ are the angular-diameter distances to the lens, the source, and between the lens and the source, and $r=\sqrt{z^2 + D_L^2\theta^2}$. We can, then, decompose the lensing potential in our model (Eq.~\eqref{eq:model}) as
\begin{equation}
    \psi = \psi_{\rm GR} + (\gamma_{\rm PN}-1) \Delta \psi,
    \label{eq:gammaPN}
\end{equation}
where $\psi_{\rm GR}$ is the usual lensing potential in GR, and 
\begin{equation}
 \Delta \psi (\theta) = 2 \int_\Lambda^\infty dr  \dfrac{r \, \Phi(r)}{\sqrt{r^2 - D_L^2\theta^2}}.
 \label{eq:PsiPN}
\end{equation}
is the correction due to screening. We will assume that the lens has a simple power-law form, as commonly done for time-delay analyses (although see the caveats in~\cite{Schneider:2013sxa,Unruh:2016adf,Munoz:2017cll}), and additionally impose spherical symmetry.
In that case, the mass density is $\rho(r) = \rho_0 (r/r_0)^{-\gamma'}$ with an index $1 < \gamma' < 3$, (not to be confused with the Post-Newtonian parameter $\gamma_{\rm PN}$), and where the $\rho_0$ and $r_0$ constants set the mass-scale for the lens. 
Then, the Newtonian potential is given by
\begin{equation}
    \Phi = \frac{4 \pi \rho_0 r_0^{\gamma'}}{(\gamma'-3)(\gamma'-2)}\,r^{2-\gamma'}.
    \label{eq:PhiGR}
\end{equation}
This yields a lensing potential in GR of~\cite{Barkana:1998qu,Suyu:2008zp,Suyu:2012rh}
\begin{equation}
\psi_{\rm GR} (\theta) = \dfrac{\theta_{E,\rm GR}^{\gamma'-1}}{3-\gamma'} ~\theta^{3-\gamma'},
\end{equation}
where the $r_0$ and $\rho_0$ parameters have been combined to obtain the Einstein angle $\theta_{E,\rm GR}$, where the subscript GR indicates that it is the value that would be inferred in GR, and acts as an overall normalization.
We can find the deflection angle simply as
\begin{equation}
\alpha_{\rm GR} (\theta) = \partial_\theta \psi_{\rm GR}(\theta) = 
\theta_{E,\rm GR}^{\gamma'-1} \theta^{2-\gamma'},
\end{equation}
where throughout we assume that $\gamma'\neq2$ for simplicity.

We can straightforwardly integrate the potential in Eq.~\eqref{eq:PsiPN} to find the PN correction to the lensing potential to be
\begin{eqnarray} \Delta\psi(\theta)&=&\dfrac{c\, \theta_{E,\rm GR}^{\gamma'-1}\, }{3-\gamma'} \left(\dfrac{D_L}{\Lambda}\right)^{\gamma'-3}
\\&~&~\times~{}_2F_1\left[\dfrac{1}{2},\dfrac{\gamma'-3}{2},\dfrac{\gamma'-1}{2};\left(\dfrac{D_L\theta}{\Lambda}\right)^2\right],\nonumber
\label{eq:deltapsi}
\end{eqnarray}
where ${}_2F_1$ is the hypergeometric function,
\begin{equation}
c = \dfrac{1 }{2\sqrt{\pi}}~ \dfrac{\Gamma\left(\dfrac{\gamma'}{2}-1\right)}{\Gamma\left(\dfrac{\gamma'-1}{2}\right)},
\label{eq:const}
\end{equation}
and $\Gamma$ is the Gamma function.
From this equation the PN correction to the deflection angle can be trivially found as $\Delta \alpha = \partial_\theta \Delta \psi$.

\section{Time Delays}
\label{sec:TimeDelays}

Our goal in this work will be to constrain deviations from GR, parametrized through $\gamma_{\rm PN}$, for different screening distances $\Lambda$.
As discussed in the introduction, we will use time-delay measurements of strongly lensed quasars to do so, as they provide us with an independent measurement of the gravitational potential at the lens.
We assume a standard flat $\Lambda$CDM cosmology, with an expansion rate today of $H_0=70$ km s$^{-1}$ Mpc$^{-1}$, and matter abundance of $\Omega_M = 0.27$, to calculate cosmological distances.

Given our lens model, which predicts the potential $\psi_i$ and deflection $\alpha_i$ at each image position $\theta_i$, we can calculate the time delay between lensed images $i$ and $j$ as~\cite{Narayan:1996ba}
\begin{equation}
    \Delta t_{ij} = \frac{1}{2}(1+z_L)\frac{D_L D_S}{D_{LS}}\left[ (\alpha_i^2 - \alpha_j^2) - (\psi_i - \psi_j)\right].
\end{equation}

We will assume that all the parameters are well known from the positions and fluxes of the images, and simply use the observed time delays $\Delta t_{ij}^{\rm obs}$ to measure $\gamma_{\rm PN}$.
There are, however, two subtleties that we have to address before being able to do so.

\subsection{Shifts in Parameters}

The correction $\Delta \psi$ to the lensing potential, from Eq.~\eqref{eq:deltapsi}, affects not only the time delay but also the image positions
for finite values of $\Lambda$ (always larger than $R_E=D_L\theta_E$, though).
In this case we cannot keep all the lens parameters fixed as we vary $\gamma_{\rm PN}$, as their best-fit values will change accordingly.
A complete way to include this effect would be through a Markov-Chain Monte Carlo (MCMC) search in parameter space, including $\gamma_{\rm PN}$ along the rest of parameters ($\theta_E, \gamma'$, etc.).
This is a costly procedure, so in this first study we will instead account for the parameter shifts by keeping the observed $\theta_{E,\rm obs}$ constant, by changing the (unobservable) GR Einstein angle $\theta_{E,\rm GR}$ as a function of $\gamma_{\rm PN}$. We achieve this by numerically solving for the Einstein angle from the expression 
\begin{equation}
\theta_{E,\rm obs} = \alpha_{\rm GR}(\theta_{E,\rm obs}) + \Delta \alpha(\theta_{E,\rm obs} ).
\end{equation}
     We have found that using the approximation that
\begin{equation}
{}_2F_1\left[\dfrac{1}{2},\dfrac{\gamma'-3}{2},\dfrac{\gamma'-1}{2};x^2 \right] ~\approx~ 1\,-\,\frac{1}{2}\,(3-\gamma')\,x^2 ,
\end{equation}
valid for $x\ll 1$,
the solution can be analytically found to be
\begin{eqnarray}
   \theta_{E,\rm GR}&=& \left[\theta_{E,\rm obs}^{1-{\gamma'}}-c\,({\gamma_{\rm PN}}-1) \left(\frac{\Lambda} {D_L}\right)^{1-{\gamma'}}\right.\cr&\qquad&\times\left.\left\{1+\frac{1}{2}(3-\gamma')\left(\frac{D_L \theta_{E,\rm obs}}{\Lambda}\right)^2\right\} \right]^{\frac{1}{1-\gamma'}}
\end{eqnarray}
to great accuracy, where $c$ is as in Eq.~\eqref{eq:const}.
This illustrates the shift to the input $\theta_{E,\rm GR}$ that has to be performed to obtain the observed $\theta_{E,\rm obs}$. However, we will use the exact numerical solution above, since results differ marginally for very small screening distances $\Lambda\sim R_E$.

\subsection{The Mass-Sheet degeneracy}

One of the main obstacles in using time-delay measurements from strong lenses is the so-called mass-sheet degeneracy (MSD), whereby a coordinate transformation in the unobservable impact parameter
\begin{equation}
 \pmb \beta \to (1-\kappa_{\rm ext}) \pmb \beta,
\end{equation}
accompanied by a shift in the lensing potential of
\begin{equation}
\psi (\theta) \to (1-\kappa_{\rm ext})\psi (\theta) + \dfrac{\kappa_{\rm ext}}{2}\, \theta^2
\end{equation}
results in the same image positions and fluxes, while shifting the time delays by a factor of $(1-\kappa_{\rm ext})$~\cite{1985ApJ...289L...1F,Saha:2000kn}.
This additional term in the lensing potential corresponds to a constant external convergence $\kappa_{\rm ext}$, which naturally appears due to mass along the line of sight to the source~\cite{Seljak:1994wa,Tihhonova:2017mym}.

There are two avenues to breaking the MSD, and both can be combined to improve the accuracy of $H_0$  measurements from strong-lensing time delays. 
The first is using simulations to obtain a probability density function (PDF) for $\kappa_{\rm ext}$, using both galaxy counts and shear information from the lens field of view~\cite{Momcheva:2005ex,Fassnacht:2009fi,Treu:2016ljm,McCully:2016yfe}. 
The second is dynamical measurements of the lens, where the velocity dispersion $\sigma_*^2$ of stars is directly related to the enclosed total (visible and dark-matter) mass~\cite{Barnabe:2007ns}.
We note that $\sigma^2_*$ is typically measured through spectroscopy at a small radius $R_{\rm eff}\sim$ kpc, where by construction our modifications to GR are screened to not alter galactic dynamics.
Even when including dynamical and simulation information, the MSD dominates the uncertainty in strong-lensing measurements of $H_0$, as the observed time delays $\Delta t_{ij}^{\rm obs}$ are usually well measured (see, however, Ref.~\cite{TieKochanek}).

We will account for the MSD by shifting the value of the observed time delay by the expected $\kappa_{\rm ext}$ of each system, which in Refs.~\cite{Suyu:2009by,Suyu:2012aa} are obtained by combining simulations with the observed external shear in each system, to obtain the component due to the lens as
\begin{equation}
\Delta t_{ij}^{\rm lens} = \dfrac{\Delta t_{ij}^{\rm obs}}{1-\kappa_{\rm ext}}.
\end{equation}
Additionally, given the small uncertainties in the observed time delays, we will assume that the time-delay error budget is dominated by the MSD, and hence the dominant parameter is the uncertainty in dynamical measurements of the lens mass. (In general, there are uncertainties related to the anisotropy in orbits that can hamper the conversion from velocity dispersion to lens mass, which we ignore.)
In that case, the uncertainty in the time delay caused by the lens is
\begin{equation}
\sigma( \Delta t_{ij}) = 2 \Delta t_{ij}^{\rm obs}\dfrac{\Delta \sigma_*}{\sigma_*},
\end{equation}
where $\Delta \sigma_*$ is the error in the velocity dispersion. 
We note that from this simple estimate we would infer a relative uncertainty in $H_0$ of $\sim 10 \%$, whereas Refs.~\cite{Suyu:2009by,Suyu:2012aa} found $\sim 7\%$ error bars per system, showing that a full analysis contains more information than our simple estimates.
We will, therefore, also show optimistic results where we assume the only source of error is the observational uncertainty in $\Delta t_{ij}$.

\section{Results}
\label{sec:Results}

We use two systems with well-measured time delays: RXJ1131-1231 and B1608+656.
Before outlining the characteristics of these two systems,
we note that there are two additional strong-lens time-delay systems employed by the H0LiCOW collaboration to measure $H_0$, HE0435-1223 and SDSS 1206+4332, which we do not analyze.
For HE0435-1223 we would have to include a nearby perturber that cannot be accounted for as an external convergence~\cite{Wong:2016dpo}, which explicitly breaks spherical symmetry.
Similarly, the SDSS 1206+4332 system cannot be well approximated through a spherical lens, as the two images (A and B) are not coaxial with the lens center~\cite{Birrer:2018vtm}.
We leave for future work improving upon our spherically symmetric lens model to be able to employ these (and other) complicated systems in our analysis.
Let us briefly describe the two systems that we study.

\begin{figure}[htbp!]
\includegraphics[width=0.49\textwidth]{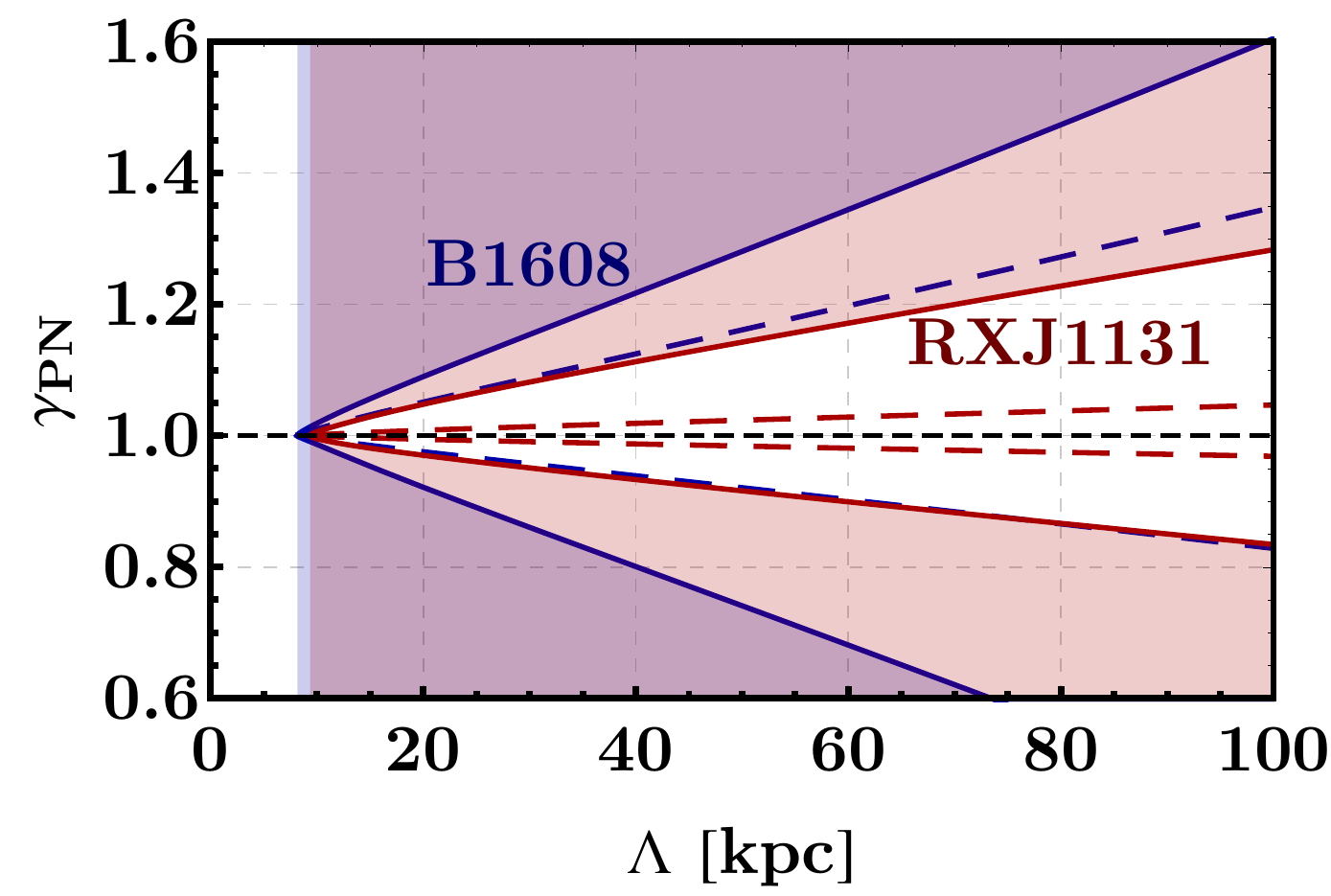}
\caption{The shaded areas are ruled out by time-delay measurements of the two systems we consider, \textcolor{red}{RXJ1131-1231} and \textcolor{blue}{B1608+656}.
The dashed lines show the constraint that could be achieved without the MSD (i.e., if the errorbars in the time delay were the limiting factor in the analysis).
The black dashed line at $\gamma_{\rm PN}=1$ represents the prediction from GR, and the slight asymmetry around this value is due to a non-unity best-fit value in our analysis.
}.
\label{fig:constraint}
\end{figure}

\subsection{RXJ1131-1231}
This system, first discovered in Ref.~\cite{Sluse:2003iy}, consists of a source QSO at $z_s=0.658$, strongly lensed by a galaxy at $z_L=0.295$. 
RXJ1131 has been carefully studied, and used to measure the Hubble expansion rate to better than $10\%$ precision~\cite{Suyu:2012aa}.
In addition, RXJ1131 has been used to set lower bounds on the mass of a putative warm-dark matter candidate through substructure constraints \cite{Birrer:2017rpp}.

As we are assuming spherical symmetry, we can only use the two QSO images that are co-axial with the lens, which for this system are images A and D, following the nomenclature in Ref.~\cite{Suyu:2012aa}.
We take the image positions from Ref.~\cite{Jee:2014uxa}, where we find their angular distances to the lens center to be $\theta\equiv\sqrt{(\pmb \theta_A-\pmb \theta_L)^2}=2.11$ arcsec, where $\pmb \theta_L$ is the centroid of the lens.
Similarly, $\theta'\equiv\sqrt{(\pmb \theta_D-\pmb \theta_L)^2}=1.12$ arcsec.
The observed time delay between A and D is $T_{ij} = -91 \pm 2.1$ days~\cite{Tewes:2012zz}, and this system has a relative error in the velocity dispersion of $\Delta \sigma_*/\sigma_* = 6\%$~\cite{Jee:2014uxa}.
Finally, the main lens has an observed Einstein angle of $\theta_{E,\rm obs} = 1.64$ arcsec, a power-law index of $\gamma'=1.95$ and an expected median value of external convergence of $\kappa_{\rm ext}=0.09$~\cite{Suyu:2012aa}.

In addition to the main lens, there is a small satellite (S), with an Einstein angle of $\theta_E^{S} = 0.2$ arcsec. As its existence breaks spherical symmetry we do not include it in our analysis, but we have checked that simply adding $\theta_E^{S}$ to the observed Einstein angle does not produce a departure from GR within our error bars.

\subsection{B1608+656}

This system, first discovered in Ref.~\cite{Myers1995}, consists of a QSO at $z_s=1.394$ and a lens at $z_L=0.6304$.
Again we only use the two coaxial images with the lens, C and D, where we read the image and lens-centroid positions from Fig.~7 of~\cite{Suyu:2008zp}, to obtain $\theta\equiv\sqrt{(\pmb \theta_C-\pmb \theta_L)^2}=1.18$ arcsec, and $\theta'\equiv\sqrt{(\pmb \theta_D-\pmb \theta_L)^2}=0.87$ arcsec.
The observed time delay between these two images is $T_{ij} = -41^{+2.5}_{-1.8}$ days~\cite{Fassnacht:2002df}

The main lens has an Einstein angle of $\theta_{E,\rm obs} = 0.924$ arcsec ~\cite{Koopmans:2003ha}, and a power-law index of $\gamma'=2.08$. 
We will also ignore the small satellite, G2, for this lens, which has an Einstein angle of $\theta_E^{G2}=0.28$ arcsec.
The median value of the external convergence for this system is $\kappa_{\rm ext}=0.08$, and the uncertainty in the velocity dispersion is $\Delta \sigma_*/\sigma_* = 6\%$~\cite{Jee:2014uxa}.


\subsection{Constraints on $\gamma_{\rm PN}$}

We use the data for the two strong lenses described above to constrain the PN parameter $\gamma_{\rm PN}$ from Eq.~\eqref{eq:gammaPN}.
We show, in Figure~\ref{fig:constraint}, the maximum and minimum values for $\gamma_{\rm PN}$ allowed by each of our strong-lensing systems at 68\% C.L., as a function of the physical cutoff scale $\Lambda$. 
We have assumed that the modeling uncertainties are dominated by the stellar dynamics, which induces relative errors of $\sim10\%$ in the time delay due to the lens itself, although we also show the result if the only uncertainty were measurement errors in $\Delta t_{ij}^{\rm obs}$ (of a few percent).
As is clear from Fig.~\ref{fig:constraint}, strong-lensing time delays constrain departures from GR for all screen radii $\Lambda$ that we study, with our constraints scaling as $|\gamma_{\rm PN}-1| \lesssim 0.2 \times  (\Lambda/100\,\rm kpc)$, for $\Lambda \geq R_E \approx 10$ kpc.
We note, in passing, that the screening radii in different theories of massive gravity are typically determined by a generalized geometric mean of the Schwarzschild radius of the mass source and the Compton wavelength of the massive graviton~\cite{Babichev:2013usa}. 
For a galaxy with mass $10^{12}\,M_\odot$ and a Hubble-radius graviton, the direct geometric mean yields a screening radius of $\sim20$ kpc, in the range that we probe.

\subsection{Discussion}

Interestingly, for small values of $\Lambda$ (but still satisfying $\Lambda \geq R_E$) the behavior of the constraints is more complicated.
In this regime, strong-lensing time delays can constrain deviations from GR at the percent level, given the large expected change in the image positions and time delays.
We note, however, that our approximation that all lens parameters are fixed might not hold for small values of $\Lambda$, so a full MCMC analysis is required to fully establish these constraints. 
Nonetheless, for $\Lambda \gtrsim 15$~kpc we have tested the sensitivity of our results to changes in parameters. We have varied the cosmological parameters $\Omega_M$ and $H_0$ within the range suggested by Planck as well as local measures of the expansion rate. We have also considered a range of values of the power-law lens profile, $\gamma'$, as indicated by the non-coaxial images. We find that our results are broadly insensitive to these changes, with the best-fit $\gamma_{\rm PN}$ shifting less than a sigma, and its error only changing at the 10\% level.

\begin{figure}[hbtp]
\includegraphics[width=0.49\textwidth]{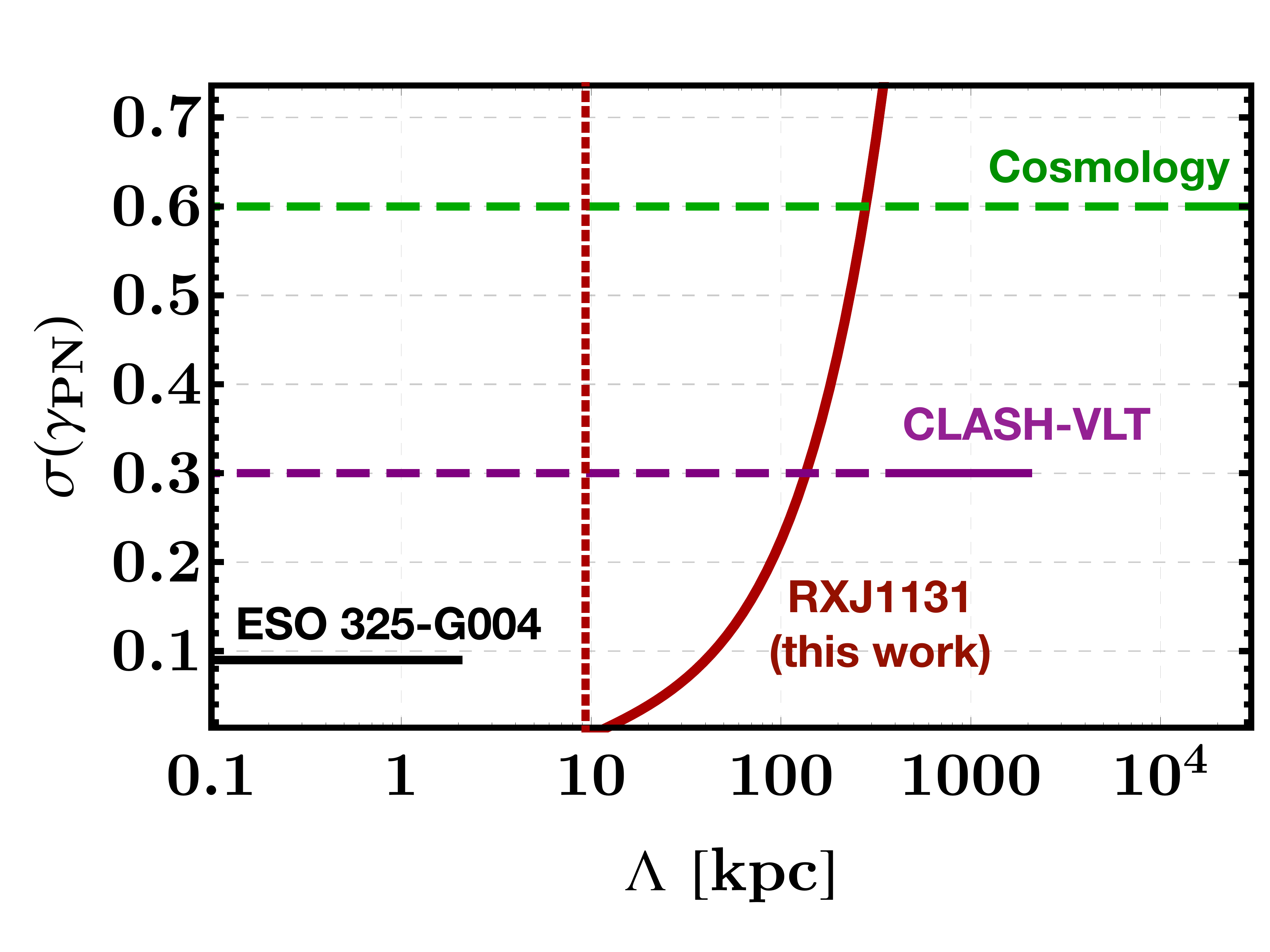}
\caption{1-$\sigma$ errors on the post-Newtonian time-slip parameter $\gamma_{\rm PN}$ as a function of the screening length $\Lambda$.
In black we show the results of Ref.~\cite{Collett:2018gpf}. Their scale-independent post-Newtonian parameter is equivalent to $\gamma_{PN}$ in our model Eq.~\eqref{eq:model} for $\Lambda\leq R_E$, where $R_E$ is the Einstein radius of the ESO 325-G004 lens, as shown. In solid purple we show  similar analysis from Ref.~\cite{Pizzuti:2016ouw} on a cluster from CLASH and VLT observations, where the cutoff scale has to be below 2 Mpc.
In solid green we show the results from cosmological observations from Ref.~\cite{Ade:2015rim}, using CMB data plus large-scale structure, which is valid for cutoffs $\Lambda<1/k_{\rm max}$, where we take the maximum wavenumber in the analyses to be the nonlinear scale $k_{\rm max} = 0.2\,\rm Mpc^{-1}$.
In solid red we show our result from strong-lensing time delays using RXJ1131-1231, which clearly places the strongest constraints on $\gamma_{\rm PN}$ for screening lengths between $R_E=10$ kpc (which we show as a vertical dotted red line) and 200 kpc. Long dashed lines are included to guide the eye.
}
\label{fig:constrainedjoint}
\end{figure}

We note that previous work has searched for modifications from GR using strongly lensed systems without observed time delays~\cite{Bolton:2006yz,Smith:2009fn,Schwab:2009nz}, constraining $\gamma_{\rm PN}$ to within 10 percent of unity using galaxies~\cite{Collett:2018gpf}, or within 30 percent using clusters~\cite{Pizzuti:2016ouw}.
These analyses, however, only search for a distance-independent deviation from GR by comparing the dynamical mass (obtained from velocity dispersions) in each lens with their Einstein Radius (which provides a measure of the lens mass).
This can be regarded as the opposite scenario $\Lambda \ll R_E$ of our parametrization in Eq.~\eqref{eq:gammaPN}.
In that case the dynamics of the lens (given by $\Phi$) and its light deflection (given by $\Phi+\Psi$) are different at all relevant scales of the problem~\cite{Smith:2009fn}.
Our work studies the complementary range $\Lambda\gtrsim R_E$, using time delays as an additional datum to measure $\gamma_{\rm PN}$, in a similar manner to $H_0$ measurements.
We compare our constraints from RXJ1131-1231 with those of previous work in Fig.~\ref{fig:constrainedjoint}, along with that obtained from CMB and large-scale structure data~\cite{Ade:2015rim}.
Clearly, strong-lensing time delays are the most sensitive probe to departures from GR in the range $\Lambda=10-200$ kpc.

We find large differences in constraining power from one system to another, as RXJ 1131 can place constraints on $\gamma_{\rm PN}$ twice as stringent as B1608.
Nonetheless, we can forecast what our constraints would be given a number $N$ of strong lenses with well-measured time delays (and mass models), by averaging the errors for the two systems that we have and dividing by $\sqrt{N}$, as different systems are uncorrelated. 
Doing so we estimate a forecasted constraint of
\begin{equation}
    \sigma(\gamma_{\rm PN}) \approx 0.07 \times (\Lambda/100\,\rm kpc) \times (N/30)^{-1/2},
\end{equation}
reaching precision below 10\%.
This is a conservative estimate, as our analysis only employs the two co-axial strongly lensed images.

For completeness, we note that it is possible that the potential $\Phi$ departs from the Newtonian form beyond the screening radius in different ways. 
For instance, in some theories the effective Newton's constant is a function of redshift~\cite{Navarro:2005gh,DeMartino:2018yqf}. 
Given that the functional form of the departure varies from one theory to another, we only present results for our pure Newtonian model (Eq.~\ref{eq:PhiGR}).
This model can represent a wide class of massive-gravity theories in the general framework of bimetric gravity \cite{Platscher:2018voh}, where the correction term to the Newtonian form can be neglected for the graviton masses that we can probe.


\section{Conclusions}
\label{sec:Conclusions}

In this work we have studied the effects of a cosmic time slip in strong-lensing time-delay measurements.
We developed a phenomenological model of gravitational screening where the {Newtonian} potential equals {the one in} GR for small radii, and transitions to a value that is $\gamma_{\rm PN}$ times  that at large distances. For computational convenience we used an abrupt transition at a screening length $\Lambda$, and calculated its effects on the lensing potential $\psi$, which we used to derive the modified Einstein angle $\theta_E$ and time delays $\Delta t$ for spherically symmetric lenses.

Using the formalism outlined above, and the data from the two strong-lens systems, B1608+656 and RXJ1131-1231, we were able to constrain deviations from GR with screening lengths $\Lambda = 10-200$ kpc at the ten percent level. These are the first constraints on modified gravity with screening in this range, and the first to make use of time delays. In this analysis we employed a simple spherical-symmetric model, and kept the observed properties of the lens fixed, only using the time delay as a datum to constrain $\gamma_{\rm PN}-1$. In future work we will use MCMCs to measure all lens parameters and $\gamma_{\rm PN}$ simultaneously, which will allow more robust and precise constraints, as well as to include the results from systems more complicated than the ones that we studied. 
Nonetheless, this first study imposes the strongest constraints to departures from GR in theories with screening scales between 10 and 200 kpc,
showing the promise of this method.

\acknowledgements

We are pleased to thank Tristan Smith and David Kaiser for helpful discussions.
This work was supported in part at Dartmouth by DOE grant DE-SC0010386, at Harvard by the DOE grant DE-SC0019018, and at Johns Hopkins by NASA Grant no.\ NNX17AK38G, NSF Grant No.\ 1818899, and the Simons Foundation.

\bibliography{biblio}

\end{document}